# Microscopic Description of $K^+$ Scattering on $^4$He , $^{16}$O and $^{40}$Ca Nuclei using Meson Exchange Theory

K. M. Hanna[1], Sh.M. Sewailem[1], A. G. Shalaby[2]
[1]Math. and Theor. Phys. Dep., NRC, Atomic Energy Authority, Cairo, Egypt,
[2]Benha university, Faculty of science, Department of Physics-Egypt

## Abstract

We have calculated the total cross section for $K^+ - ^4He$, $K^+ - ^{16}O$ and $K^+ - ^{40}Ca$, interactions at incident momenta of the kaon $P_{lab} < 1\,GeV\ /c$. We derived the $K^+$-nucleon optical potential according to the exchange of 3- mesons ($\sigma, \rho, \omega$) and also for 4-mesons ($\sigma, \rho, \omega, \sigma_o$) exchanged between the reactants. We showed both of the radial behavior of the real and the imaginary parts of the derived potential. Comparisons between the available experimental data, other theoretical work and the calculated total cross sections for the three studied nuclei which have shown a reasonable agreement. The extended four mesons exchanged optical potential gave better close results to the experimental data. Further, ratios of the total cross sections of the studied nuclei with respect to the total cross section with the deuteron nucleus for the two applied optical potentials were given. In addition, the ratio of the theoretical result of the total cross section for the interaction of the $K^+$ meson with the deuteron was compared with the corresponding experimental one to evaluate our theoretical results in a more clear manner.

## I - Introduction

For many years ago the $K^+$-nucleus ($K^+ A$) interaction has been studied theoretically and experimentally [1-6]. These interactions belong to strong interaction set, however the ($K^+ A$) interaction has weak character. Although the optical potential can describe the scattering process of these interactions at intermediate energies, still found theoretical unreasonable predicted results for the total cross sections at beam momenta 500-1000 MeV/c, e.g see [7-10]. The positive kaon is a pseudoscalar (odd parity, spin zero) meson. The kaon is a strange particle so that, it transfers to the nucleus an additional quantum number, i.e. strangeness S.

Investigation of the properties of the nuclear environment led to many studies to explore the properties of $\rho, \omega, \sigma$ exchange mesons in the interactions [11-15]. As the ($K^+ A$) interaction is so weak, $K^+$meson can penetrate deeper into the nucleus than in the case of nucleon- nucleon interaction.

The outline of the paper is as follows. In Section II, we give a brief review of the derivation of the effective interaction. In Section III, we present the kaon-nucleon optical potential. Section IV, contains the meson wave function and the parameters used. Section V, represents the results obtained using the derived potential and the calculated total cross section in comparison with the experimental data.

**Emails**
1- sh_m_sw@yahoo.com
2- asmaa.shalaby@fsc.bu.edu.eg

## II - Theoretical Scheme

The most important exchanged mesons in the $K^+ N$ interaction are the vector-isovector $\rho(1^-, 1)$, the vector-isoscalar $\omega$ $(1^-, 0)$ and the scalar-isoscalar $\sigma(0^+, 0)$ mesons. According to the interplay between the repulsive $\omega$ and the attractive $\sigma$ mesons, the interaction that responsible for the cancellation usually happened between these two fields.

Then the $K^+ N$ interaction potential according to the One-Boson-Exchange (OBE) model [16-19] in which $P_{lab}$ < 1 GeV/c, $V_{K^+-N}(r)$ can be written as,

$$V_{K^+N}(r) = V_\sigma(r) + V_\rho(r) + V_\omega(r) \quad (1)$$

And

$$V_\sigma(r) = -\gamma_1^0 \gamma_2^0 J_\sigma(r), \; V_\rho(r) = \gamma_1^0 \gamma_2^0 \gamma_1^\mu \gamma_2^\mu J_\rho(r), \; V_\omega(r) = \gamma_1^0 \gamma_2^0 \gamma_1^\mu \gamma_2^\mu J_\omega(r) \quad (2)$$

Where $\gamma_i^0$ and $\gamma_i^\mu$ ($i$ =1, 2) are the Dirac matrices. The J's functions, $J_\sigma(r)$ , $J_\rho(r)$ and $J_\omega(r)$ are suitable Yukawa type functions.

For better description of the theoretical results and towards a reasonable fit with the experimental data, one needs to add an additional repulsive meson in the interaction more than obtained by the $\omega$ meson, a phenomenological repulsive $\sigma_0$ -meson (which has shorter range and higher mass than the $\omega$ meson). Then the $K^+ N$ potential will be modified to the form;

$$V_{K^+N}(r) = V_\sigma(r) + V_\rho(r) + V_\omega(r) + V_{\sigma_0}(r) \quad (3)$$

Where the $\sigma_0$ structure is taken as the structure of the $\sigma$ -meson, but with opposite sign and heavier exchange mass [20] as follows:

$$V_{\sigma_0}(r) = \gamma_1^0 \; \gamma_2^0 \; J_{\sigma_0}(r) \quad (4)$$

### II.1 - Normalization of Nucleon Wave Functions

In Dirac space the normalization condition for the nucleon wave functions $f_\gamma(r)$ can be written as follows:

$$\left\langle f_\gamma(\vec{r}) \middle| f_\gamma(\vec{r}) \right\rangle = \left\langle \varphi_\gamma(\vec{r}) \middle| \varphi_\gamma(\vec{r}) \right\rangle + \left\langle \chi_\gamma(\vec{r}) \middle| \chi_\gamma(\vec{r}) \right\rangle \quad (5)$$

Where, $\varphi_\gamma(r)$ and $\chi_\gamma(r)$ are the large and small wave function components respectively.
Consequently, the normalized nucleon wave function can be expressed in the form [21, 22]:

$$\left|\varphi'(\vec{r})\right\rangle = \sqrt{1+p^2/4m_2^2c^2}\left|\varphi(\vec{r})\right\rangle \tag{6}$$

Where, $m_2$ is the mass of the nucleon and $P$ its relative momentum. One can take approximately only the first term in the expansion of the small wave function component $\chi_\gamma(\vec{r})$ in terms of the large one $\varphi_\gamma(\vec{r})$, which given by the relation [23],

$$\chi_i(\vec{r}) \simeq \frac{\vec{\sigma}_i \cdot \vec{p}}{2mc}\varphi_i(\vec{r}) \tag{7}$$

Where $\vec{\sigma}_i$ are the Pauli spin matrices.

### II.2 - Kaon–Nucleon kinematics and the Wave Function Expansion

In the $K^+N$ system in which, $m_1$ and $\vec{r}_1$ are the kaon mass and coordinate respectively, while $m_2$ and $\vec{r}_2$ are the nucleon mass and coordinate. The potential between the two inequal mass particles which we intend to take as harmonic oscillator, which enables us to deal with the two-body wave function as a product of a relative part and center of mass part coordinate wave functions [24]. Hence, the relative and C.M. coordinate values are written as follows:

$$\vec{r} = \sqrt{2}\,\frac{m_1\vec{r}_1 - m_2\vec{r}_2}{m_1+m_2}\ ; \quad \vec{R} = \sqrt{2}\,\frac{m_1\vec{r}_1 + m_2\vec{r}_2}{m_1+m_2} \tag{8}$$

Similarly, the relative momenta $\vec{P}_{kN} \equiv \vec{P}_r$ and C.M. momenta $\vec{P}_R$ are given by;

$$\vec{P}_r = \frac{m_2\vec{p}_1 - m_1\vec{p}_2}{m_1+m_2}\ ; \quad \vec{P}_R = \vec{p}_1 + \vec{p}_2 \tag{9}$$

The kaon wave function $\varphi_\alpha(r)$ is given by;

$$\varphi_\alpha(\vec{r}) = \sum_{m_{\ell\alpha}} \left(\ell_\alpha\, 0\, m_{\ell_\alpha}\, 0 \,\middle|\, \ell_\alpha\, m_{\ell_\alpha}\right) \varphi_{n_\alpha \ell_\alpha m_{\ell_\alpha}}(\vec{r}) \hat{P}_{T_\alpha} \tag{10}$$

Where $\alpha$ represents the quantum number collection in the Clebsch-Gordon coefficient ($n_\alpha, l_\alpha, j_\alpha, m_\alpha$). It is remarkable that, there is no dependence on a spin function of the projectile (kaon) as it is a spinless particle. For the expansion of the nucleon wave function $\varphi_\gamma(\vec{r})$ in terms of the spin function $\chi^{1/2}_{m_{s_\gamma}}$ and the isotopic spin $\hat{P}_{T_\alpha}$ we have:

$$\varphi_{\gamma}(\vec{r}) = \sum_{m_{\ell_\gamma}, m_{s_\gamma}} \left( \ell_\gamma \; s_\gamma \; m_{\ell_\gamma} m_{s_\gamma} \middle| J_\gamma \; m_{J_\gamma} \right) \varphi_{n_\gamma \ell_\gamma m_{\ell_\gamma}}(\vec{r}) \; \chi^{1/2}_{m_{s_\gamma}} \; \hat{P}_{T_\gamma} \tag{11}$$

Consequently, the $K^+ N$ wave function can be rewritten as follows:

$$\left\langle \varphi_\alpha(\vec{r}_1) \varphi_\gamma(\vec{r}_2) \right| = \sum_{m_{\ell\alpha} m_{\ell_\gamma}, m_{s_\gamma}} \left( \ell_\alpha, 0, m_{\ell_\alpha}, 0 \middle| J_\alpha, m_{J_\alpha} \right) \left( \ell_\gamma, \frac{1}{2}, m_{\ell_\gamma}, m_{s_\gamma} \middle| J_\gamma, m_{J_\gamma} \right) \left\langle \varphi_{n_\alpha \ell_\alpha m_{\ell_\alpha}}(\vec{r}_1) \, \varphi_{n_\gamma \ell_\gamma m_{\ell_\gamma}}(\vec{r}_2) \, \chi^{1/2}_{m_{s_\gamma}} \hat{P}_{T_\alpha} \hat{P}_{T_\gamma} \right| \tag{12}$$

The two angular wave functions for the kaon and nucleon are coupled to produce the collective wave function of the two particles as:

$$\left\langle \varphi_{n_\alpha \ell_\alpha m_{\ell_\alpha}}(\vec{r}_1) \varphi_{n_\gamma \ell_\gamma m_{\ell_\gamma}}(\vec{r}_2) \right| = \sum_{\lambda\mu} \left( \ell_\alpha \ell_\gamma m_{\ell_\alpha} m_{\ell_\gamma} \middle| \lambda\mu \right) \left\langle \varphi_{n_\alpha \ell_\alpha \, n_\gamma \ell_\gamma \lambda\mu}(\vec{r}_1, \vec{r}_2) \right| \tag{13}$$

If one considers the motion of the $K^+ N$ system in the form of the harmonic oscillator, the generalized Talmi-Moshinsky-Smirnov (GTMS) brackets for particles having different masses will be imported [25],

$$\left\langle \varphi_{n_\alpha \ell_\alpha n_\gamma \ell_\gamma \lambda\mu}(\vec{r}_1, \vec{r}_2) \right| = \sum_{n\ell NL} (n_\alpha \ell_\alpha n_\gamma \ell_\gamma \lambda) \,|\, NLn\ell\lambda) \left\langle \varphi_{NLn\ell\lambda\mu}(\vec{R}, \vec{r}) \right| \tag{14}$$

Consequently, the wave function of $K^+ N$ system can be separated into two components, the wave function of the relative motion $\vec{r}$ and the motion represents the C.M. $\vec{R}$ as follows:

$$\left\langle \varphi_{NLn\ell\lambda\mu}(\vec{R}, \vec{r}) \right| = \sum_{Mm} (L \ell M m | \lambda\mu) \left\langle \varphi_{NLM}(\vec{R}) \right| \left\langle \varphi_{n\ell m}(\vec{r}) \right| \tag{15}$$

Moreover, the spin and isotopic spin nucleon wave functions can be expanded as follows:

$$\left\langle \hat{P}_{T_\alpha} \chi^{1/2}_{m_{s_\gamma}}(2) \hat{P}_{T_\gamma} \right| = \sum_{s m_{s_\gamma} T M_T} (0 \, \frac{1}{2} \, 0 m_{s_\gamma} \,|\, s \, m_{s_\gamma})(\frac{1}{2} \, \frac{1}{2} \, T_\alpha T_\gamma T \, M_T) \left\langle \chi^{s}_{m_{s_\gamma}}(2) \hat{P}_T(1,2) \right| \tag{16}$$

The radial wave function of the relative motion of the $K^+N$ system $R_{n\ell}\left(\frac{r}{b}\right)$, can be given in the form of Laugurre polynomial $L_n^{\ell+\frac{1}{2}}$.

$$R_{n\ell}\left(\frac{\vec{r}}{b}\right)=\left[\frac{2(n!)}{\Gamma(n+\ell+3/2)}\right]^{1/2}\left(\frac{1}{b}\right)^{3/2}\left(\frac{\vec{r}}{b}\right)^{\ell}\exp\left(-\frac{1}{2}\left(\frac{\vec{r}}{b}\right)^{2}\right)L_{n}^{\ell+1/2}\left(\frac{\vec{r}}{b}\right)^{2} \tag{17}$$

with size parameters defined in relative and C.M. systems as follows.

$$b_r=\sqrt{\frac{\hbar(m_1+m_2)}{m_1 m_2\omega}}\,,\quad b_R=\sqrt{\frac{\hbar}{(m_1+m_2)\omega}} \tag{18}$$

### III - Kaon-Nucleon and Nucleus Potentials

After some arrangements, we can write the optical potential of the $K^+N$ interaction in the form:

$$\begin{aligned}V_{K^+N}(r)=&V_a(r)+2\mu\nu Q_3(2n+\ell+\frac{3}{2})\hbar\omega V_c(r)-\nu Q_3\mu^2\omega^2r^2V_c(r)-\nu Q_3\hbar^2\frac{dV_c(r)}{dr}\frac{d}{dr}\\&+\hbar^2\nu Q_3(j(j+1)-\ell(\ell+1)-s(s+1))\frac{1}{r}\frac{dV_c(r)}{dr}+\nu Q^2P_1^2V_c(r)-\\&i\ \left[\nu QP_1\hbar\frac{dV_c(r)}{dr}+\nu QP_1\hbar V_c(r)\frac{d}{dr}+\nu Q^2P_1\hbar V_c(r)\frac{d}{dr}+\nu Q^2P_1\hbar\frac{dV_c(r)}{dr}\right]\end{aligned} \tag{19}$$

where, the reduced mass, $\mu=\frac{m_1m_2}{m_1+m_2}$, $\nu=\frac{1}{4m_2^2c^2}$, $Q_3=\frac{m_2}{m_1}$ and the $\hbar\omega$ is the separation energy parameter [26] is given, phenomenologically, by

$$\hbar\omega=1.85+\frac{35.5}{A^{1/3}}$$

and

$$\begin{aligned}V_a(r)&=-V_\sigma(r)+V_\rho(r)+V_\omega(r)+V_{\sigma_0}(r)\\V_c(r)&=V_\sigma(r)+V_\rho(r)+V_\omega(r)-V_{\sigma_0}(r)\end{aligned} \tag{20}$$

To calculate the many body interacting potential $V_{K^+A}(r)$, we use, for simplicity, its approximated form as the sum of the two body $K^+N$ potential $V_{K^+N}(r)$ as follows:

$$V_{K^+A}(r)\simeq\sum_{N_i=1}^{A}V_{K^+N_i}(r) \tag{21}$$

The relation between the reduced potential and the interacting potential [27] is:

$$U_{K^+A}(r) = \frac{2\mu}{\hbar^2} V_{K^+A}(r) \tag{22}$$

## IV - The Exchange Meson Wave Function and its Parameters

In the present work, we have adopted the associated generalized Yukawa (GY) meson function [26] which is given by,

$$J_i(r) = g_i^2 \hbar c [\frac{\exp(-u_i r)}{r} - \frac{\exp(-\Lambda_i r)}{r}(1 + \frac{\Lambda_i^2 - u_i^2}{2\Lambda_i})] \tag{23}$$

Where the parameter $u_i = \frac{m_i c^2}{\hbar c}$ is associated with the masses of the exchanged mesons $m_i$, $g_i$ its coupling constants, $\Lambda_i = \frac{\lambda_i c^2}{\hbar c}$ is associated with the cutoff masses $\lambda_i$ and i stands for the exchanged particles $\sigma, \rho, \omega, \sigma_0$, to represent the static meson function. In our study for the interaction of the kaon with nucleons and nuclei, we have taken two different sets of parameters suggested by the Julich group [20, 28]. The table represents the different kinds of mesons that exchange during the interaction, $\sigma, \rho, \omega$ and the additional supposed meson $\sigma_0$. The kaon and nucleon masses are taken, respectively, as $m_1$ = 495.82 MeV/c$^2$, and $m_2$ = 938.926 MeV/c$^2$.

**Table: The meson masses, coupling constants and the cut-off parameters**

| meson | $m_i$ MeV/c$^2$ | $g_i / \sqrt{4\pi}$ | | $\lambda_i$ GeV/c$^2$ | |
|---|---|---|---|---|---|
| | | [20] | [28] | [20] | [28] |
| $\sigma$ | 600 | 2.385 | 1.300 | 1.7 | 1.7 |
| $\rho$ | 769 | 0.917 | 0.773 | 1.4 | 1.6 |
| $\omega$ | 782.6 | 2.750 | 2.318 | 1.5 | 1.5 |
| $\sigma_0$ | 1200 | 3.937 | 4.0 | 2 | 1.5 |

## V. Results and Discussions

In Figs. (1, 2) we presented our calculations for the $K^+$ - nucleon real and imaginary parts of our two constructed optical potentials based on the exchange of 3- and 4-mesons between the kaon and the nucleon. We notice that the real parts of the potentials with protons in the case of $K^+$ - proton gives maximum value at a distance of 0.5 Fermi 8.5 MeV in case of the 4M exchange potential and for $K^+$-proton about 5.5 MeV in 3M exchange model. With respect to the imaginary parts, the absorption reaches at the same distance, to about -2.5 MeV respectively. The corresponding values in the case of $K^+$- neutron were for the real parts 13 MeV (for the 4M model) and about 8 MeV (for the 3M model).

About the imaginary parts, the absorption in the 4M model, at a distance of 0.5 Fermi, was about -3.75 MeV while for the 3M model was -2.75 MeV. Moreover, we notice that all the above values tend to increase the real parts (in both optical models) and decrease the imaginary parts with decreasing the interacting energy between the particles.

Figs. 3,4 and 5 (a) show the real parts of the potentials, at an interacting energy 450 MeV, increase as the mass number of the nucleus increases while the imaginary parts, figs. 3, 4 and 5 (b) decrease with the increase of the mass number. The behavior of the imaginary parts of the potentials in case of the scattering of $K^+$ - meson with $^{40}$Ca- nucleus is quite interesting. Both the potentials begin as attractive close to center of coordinates (0.1 Fermi) with values about -80 MeV, and then they transfer to the repulsive character at a distance about 0.85 Fermi with value 55 MeV for the 3M model. The same character preserves for the 4M model with attractive part of the value -170 MeV at a distance about 0.15 Fermi, and a repulsion of the value 110 MeV.

The total cross section has been plotted versus the momentum in case of $K^+p$ using different sets of parameters compared and for both cases 3M- exchange and 4M- exchange, with the available experimental data [1] as shown in figures (7, 15). It is clear that the total cross section has shown a reasonable fit with the experimental data for each different set of parameters, however the theoretical results calculated by using parameters [28] are slightly give values better coincidence with experimental data than the results calculated using parameters [20]. Also we can see that, by adding the fourth meson ($\sigma_0$) the theoretical calculations give a better fitting with the experimental data [7] in the case of $K^+$D as shown in figs. (8, 16).

On the other hand in case of $K^+ -^4 He$ reaction, the theoretical results calculated by using parameters [28] as shown, slightly higher values more close to the theoretical calculations in [29] than that calculated using parameters in [20] and by the two different mechanisms as shown in figures (9, 17). The same case is repeated for $K^+$ -$^{16}$O and $K^+$ -$^{40}$Ca as shown in figs. (17, 19; 13, 21) ,in which the theoretical results are compared to the theoretical calculations in ref. [29] and the experimental refs. [7, 8, 30] respectively.

Further, we calculate the ratios of the total cross sections of the three studied nuclei to the total cross section of the deuteron as shown in Figs. (10, 12, 14, 18, 20, 22). From these ratios we can evidently see some EMC-effect in such soft nuclear reactions .

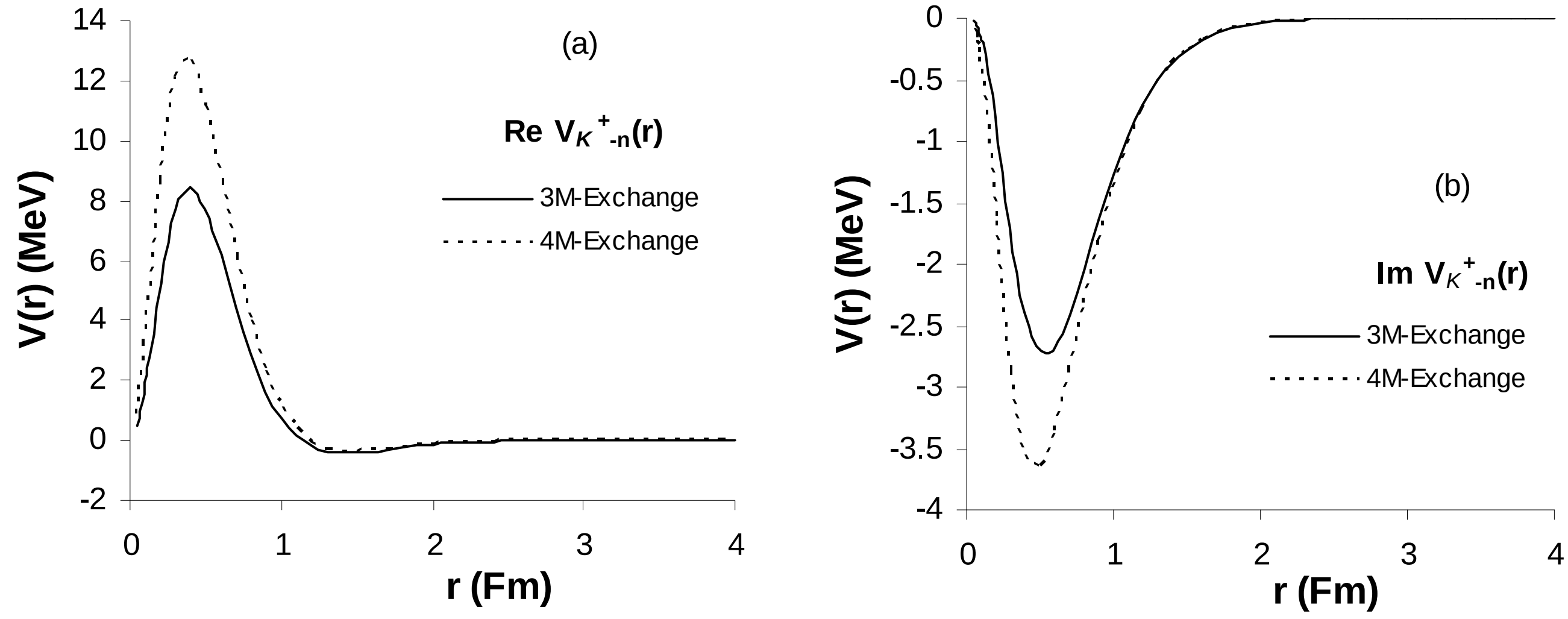

**Fig.1(a, b). The real and imaginary parts of the optical potential for 3, and 4 – exchanged mesons in case of $K^{+}n$ interaction at $P_{lab}$=450 MeV.**

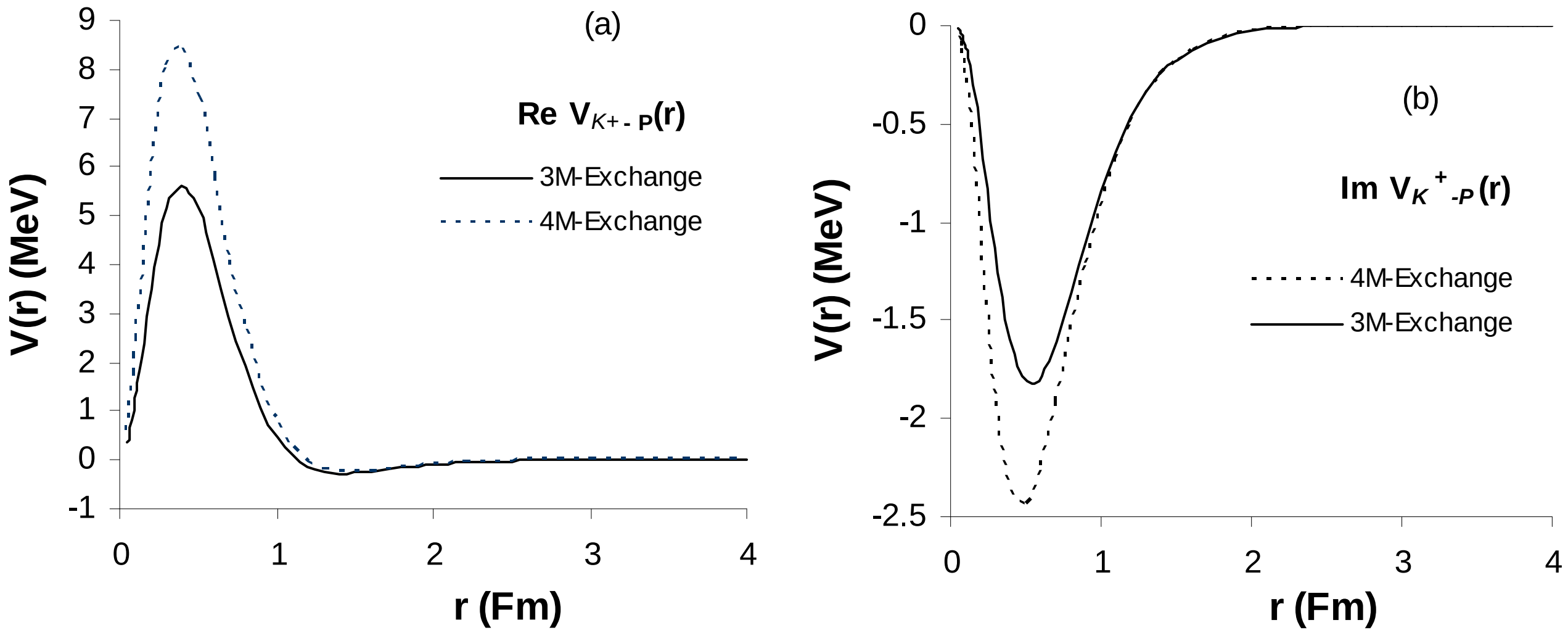

**Fig.2(a, b). The real and imaginary parts of the optical potential for 3-and 4 – exchanged mesons in case of $K^{+}P$ interaction at $P_{lab}$=450 MeV.**

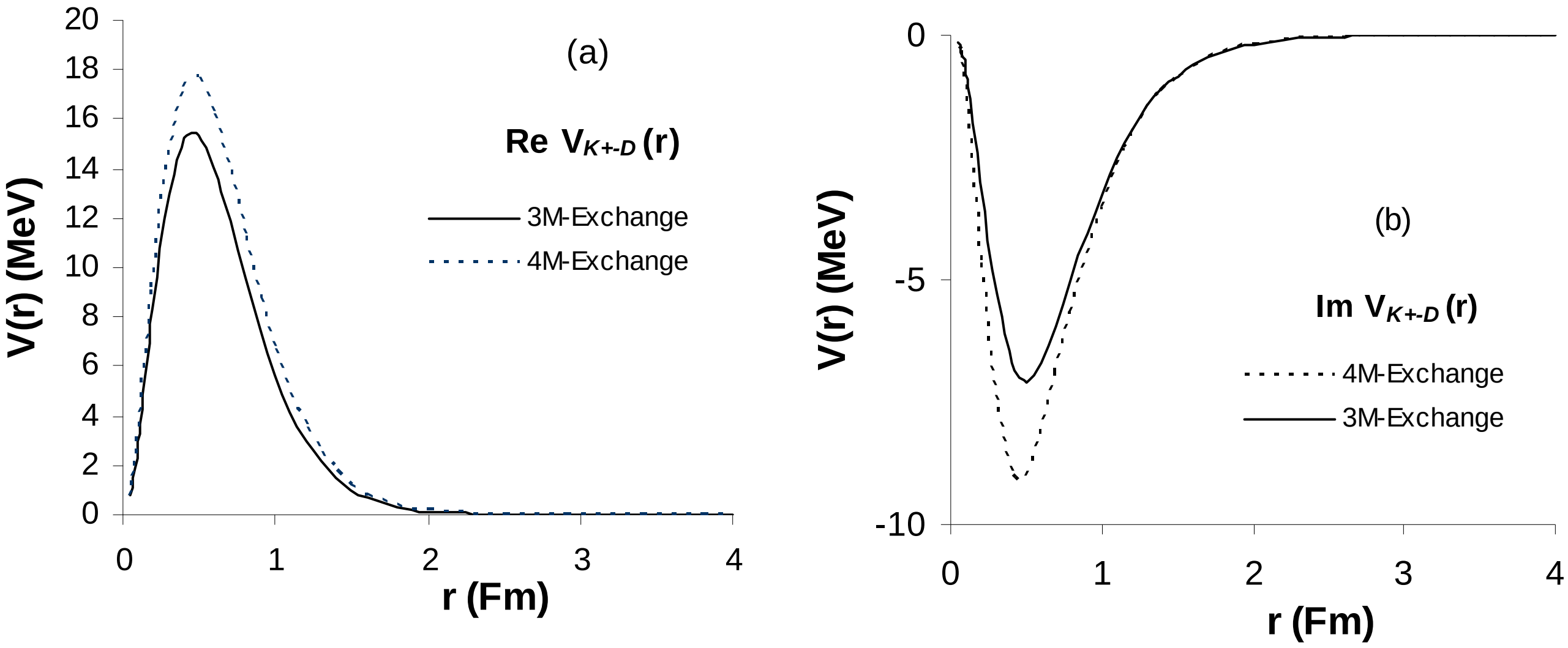

**Fig. 3(a, b). The real and imaginary parts of the optical potential for 3 and 4 – exchanged mesons in case of $K^+ - Deuteron$ interaction at P$_{lab}$=450 MeV.**

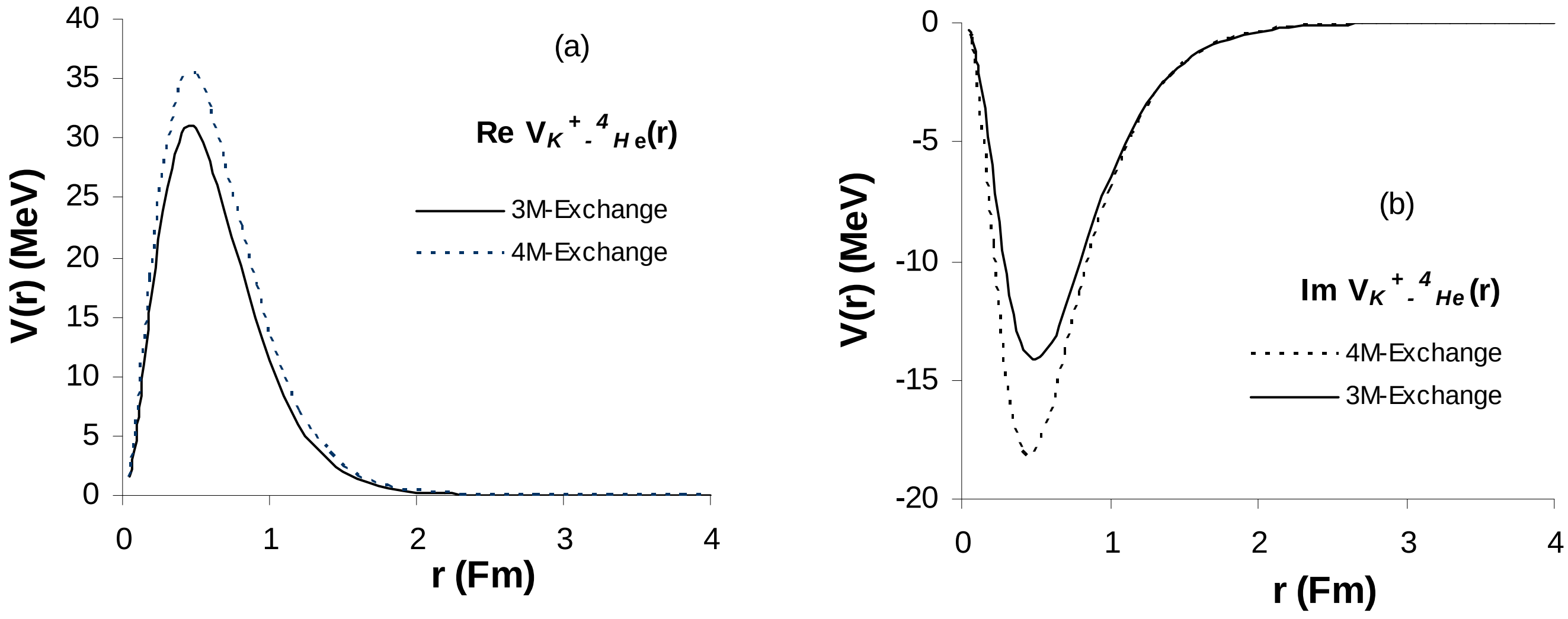

**Fig. 4(a, b). The real and imaginary parts of the optical potential for 3 and 4 – exchanged mesons in case of $K^+ - {}^4He$ interaction at P$_{lab}$=450 MeV.**

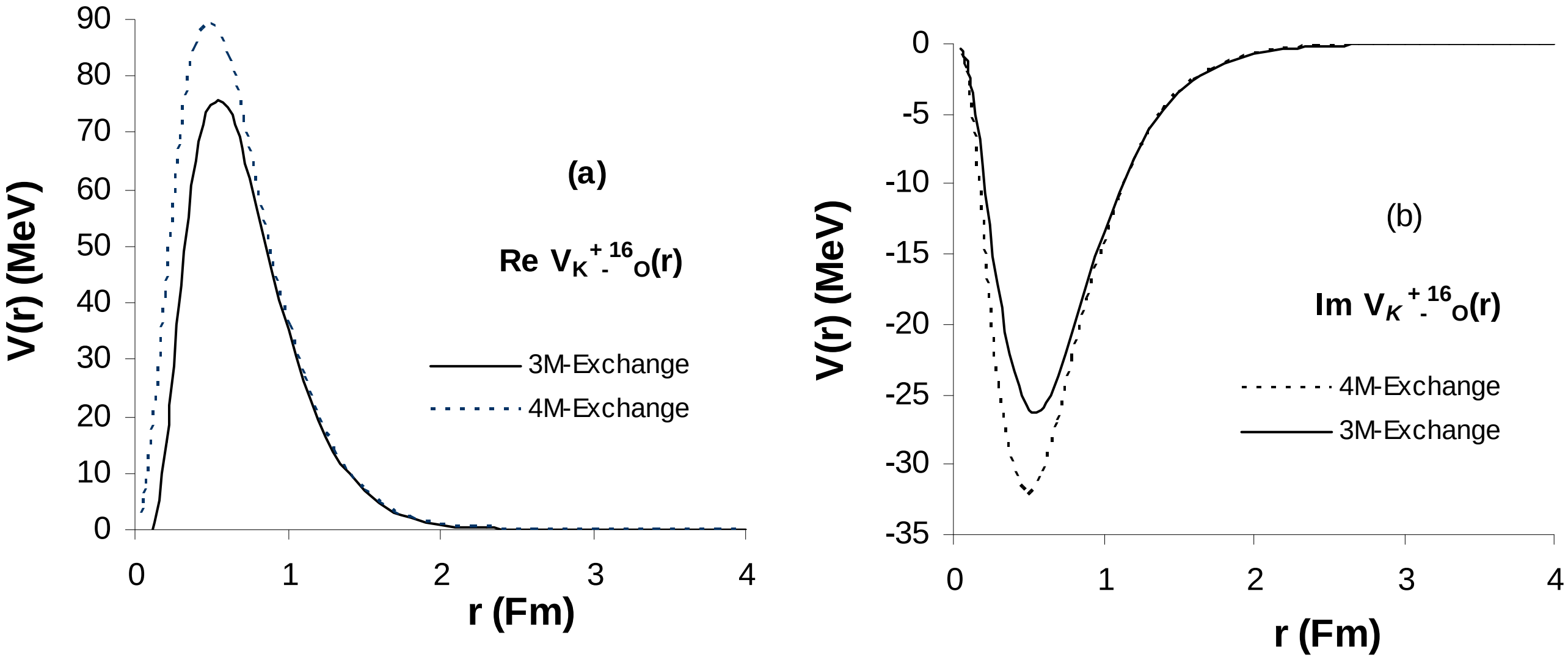

**Fig.5 (a, b). The real and imaginary parts of the optical potential for 3 and 4 – exchanged mesons in case of $K^+ -^{16} O$ interaction at $P_{lab}$=450 MeV.**

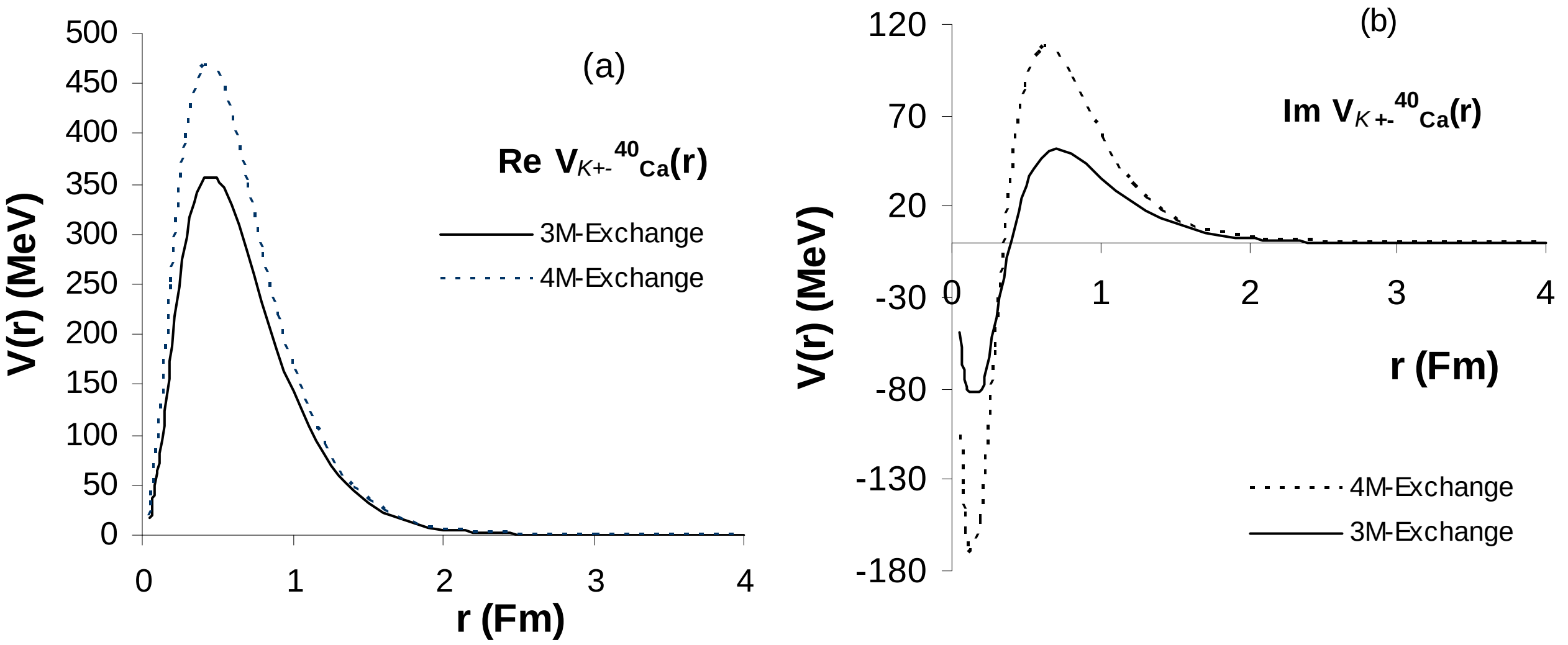

**Fig.6 (a, b). The real and imaginary parts of the optical potential for 3 and 4 – exchanged mesons in case of $K^+ -^{40} Ca$ interaction at $P_{lab}$=656 MeV.**

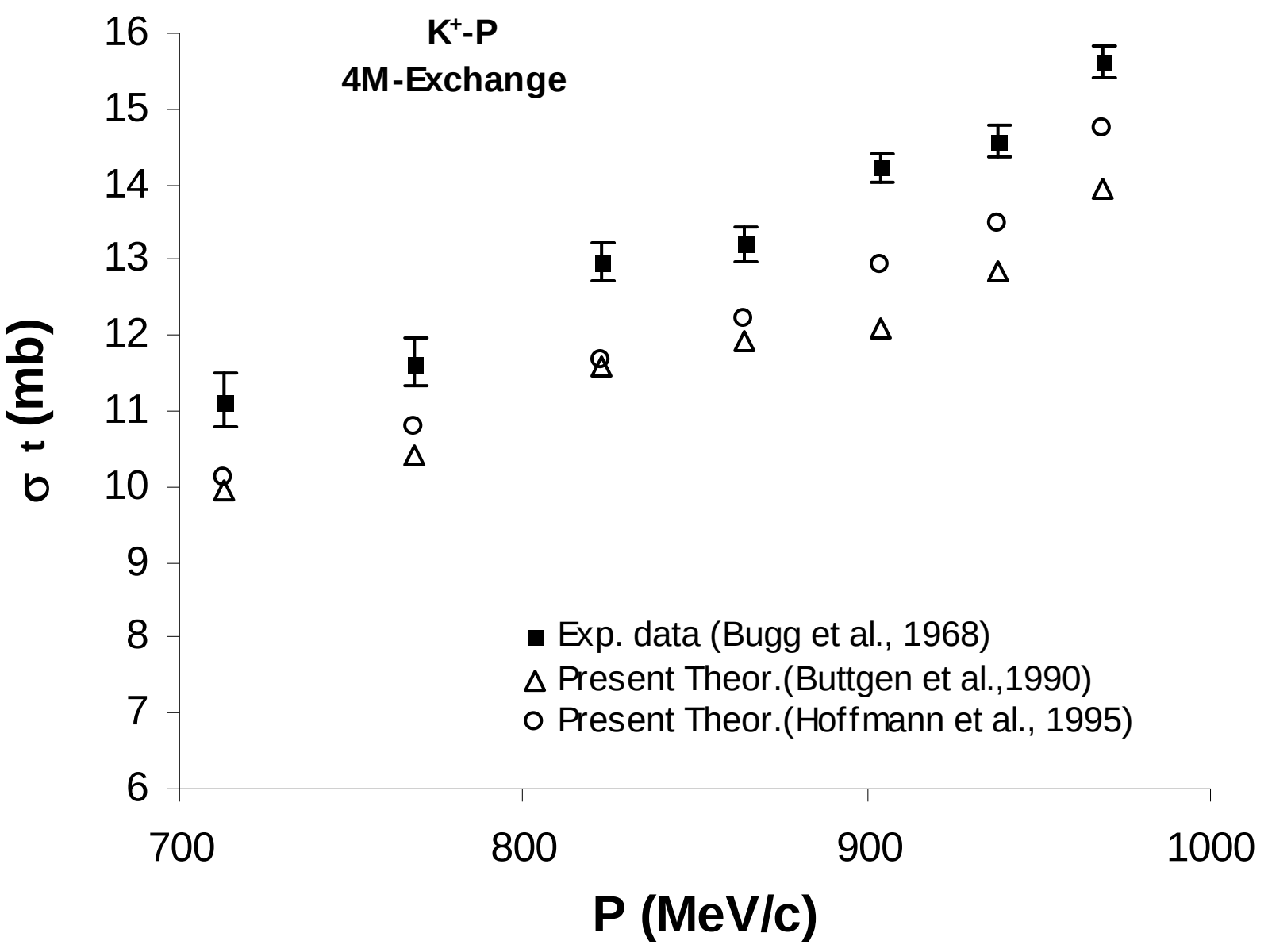

**Fig. (7) The total cross section versus the momentum in $K^+ - p$ reaction. The theoretical calculations by different parameters used as Buttgen et al. [20] and Hoffmann et al.,[28] respectively.**

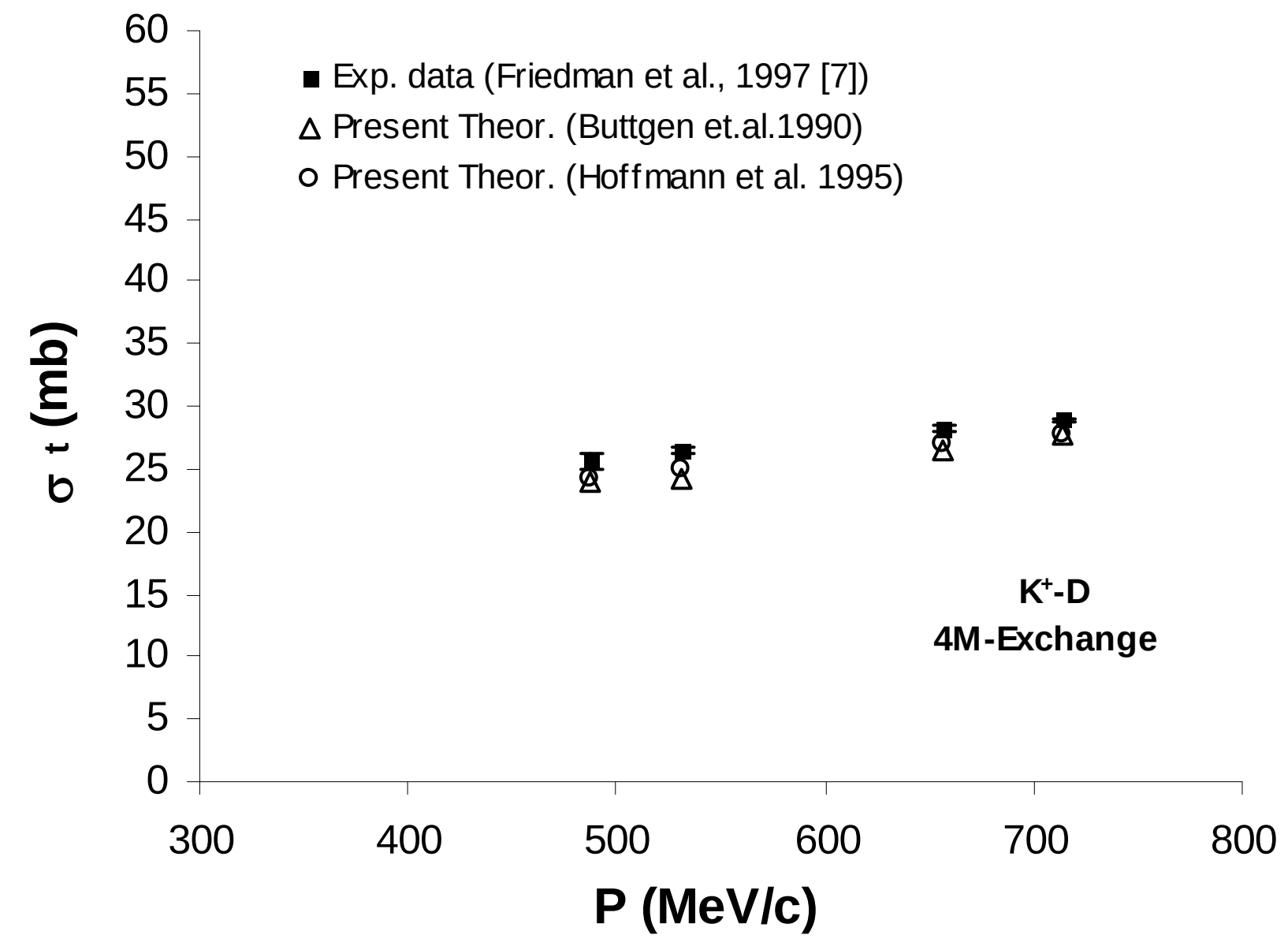

**Fig. (8) The total cross section versus the momentum in $K^+ - Deuteron$ reaction. The theoretical calculations by different parameters used as Buttgen et al. [20] and Hoffmann et al.,[28] respectively.**

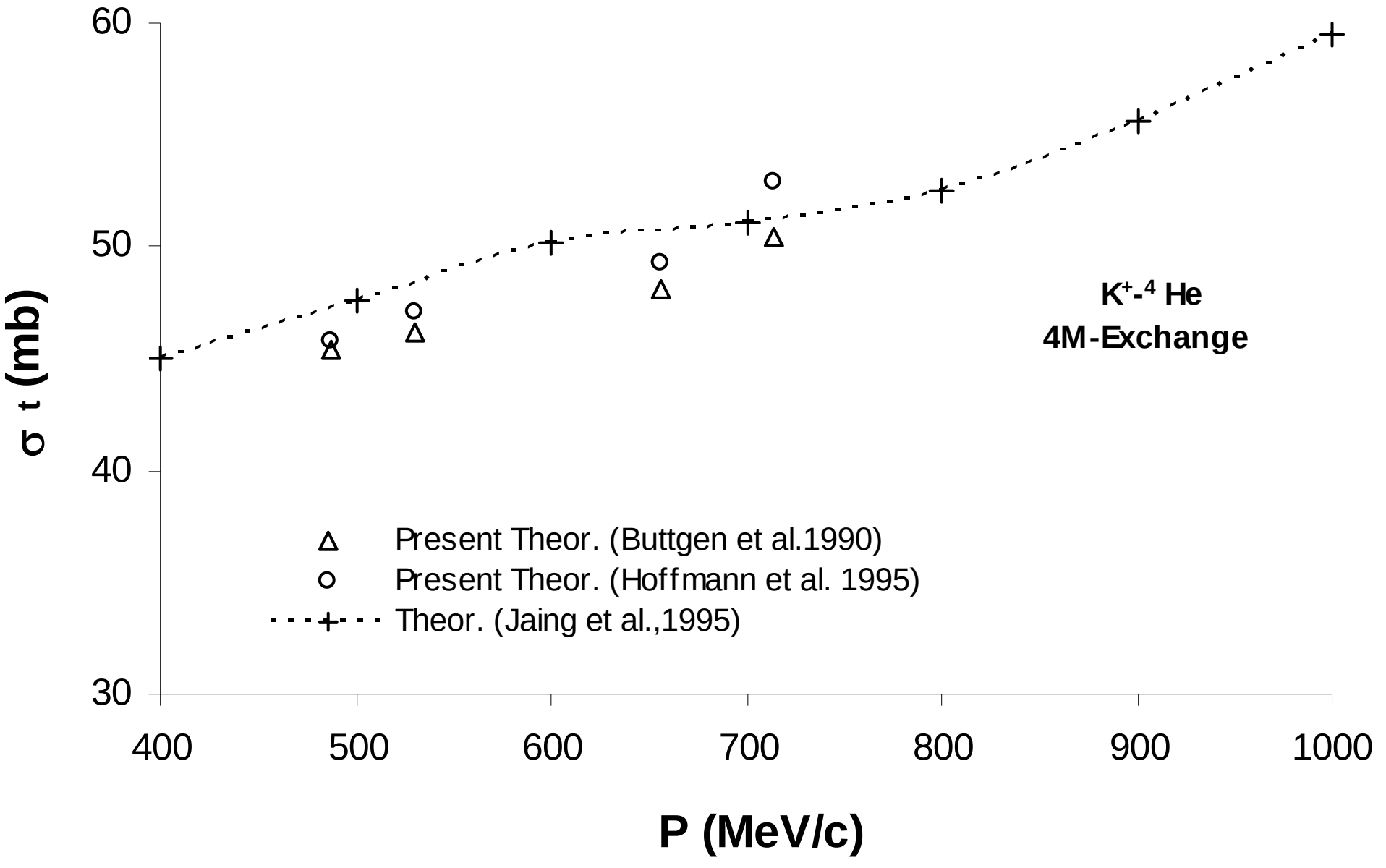

**Fig. (9) The total cross section versus the momentum in $K^{+} -^{4} He$ reaction. The theoretical calculations by different parameters used as Buttgen et al. [20] and Hoffmann et al.,[28] respectively.**

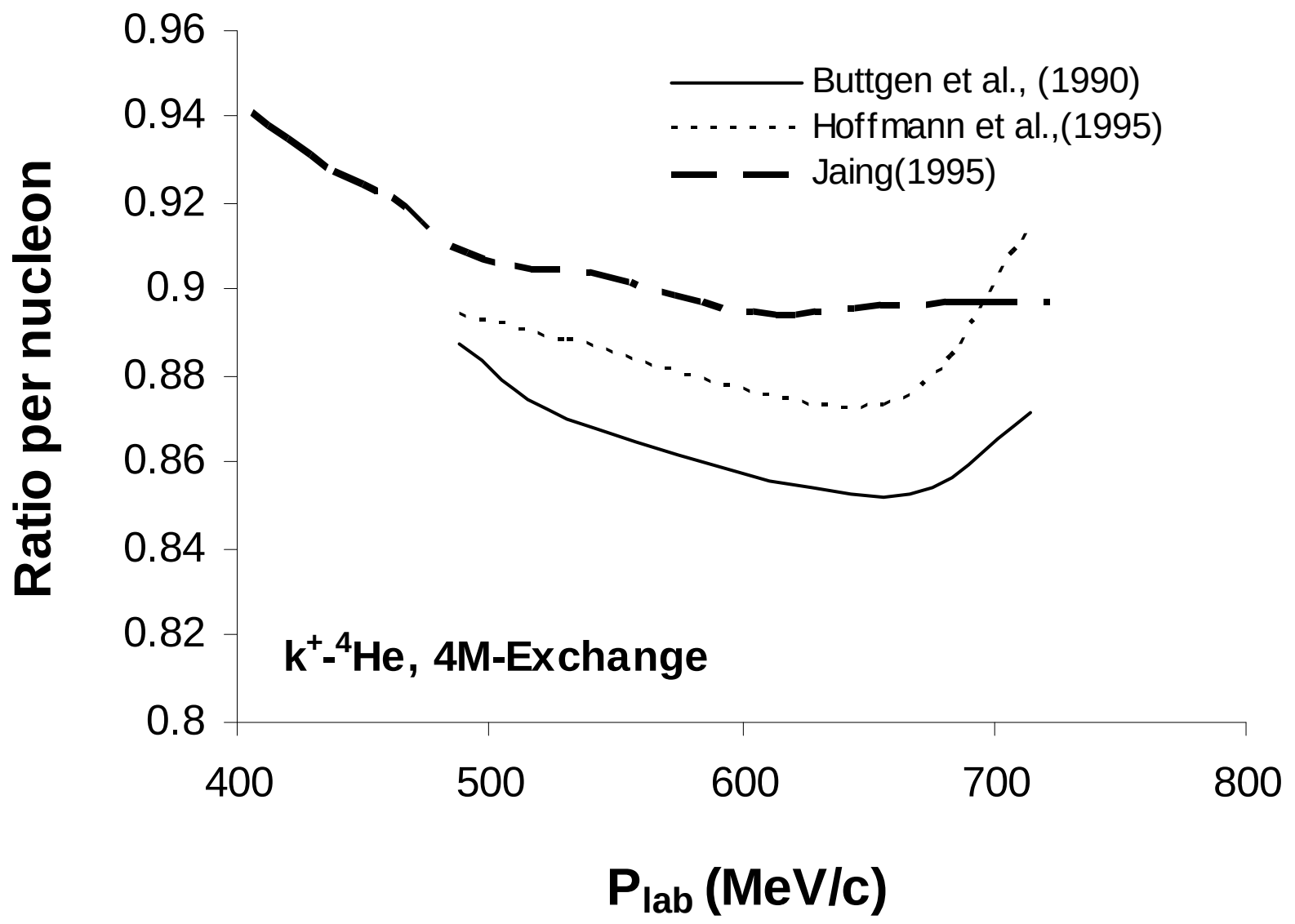

**Fig. (10) The Ratio of Helium to Deuterium cross sections per nucleon. The solid line and the dashed line indicate the theoretical calculations by different parameters used as Buttgen et al. [20] and Hoffmann et al.,[28] respectively.**

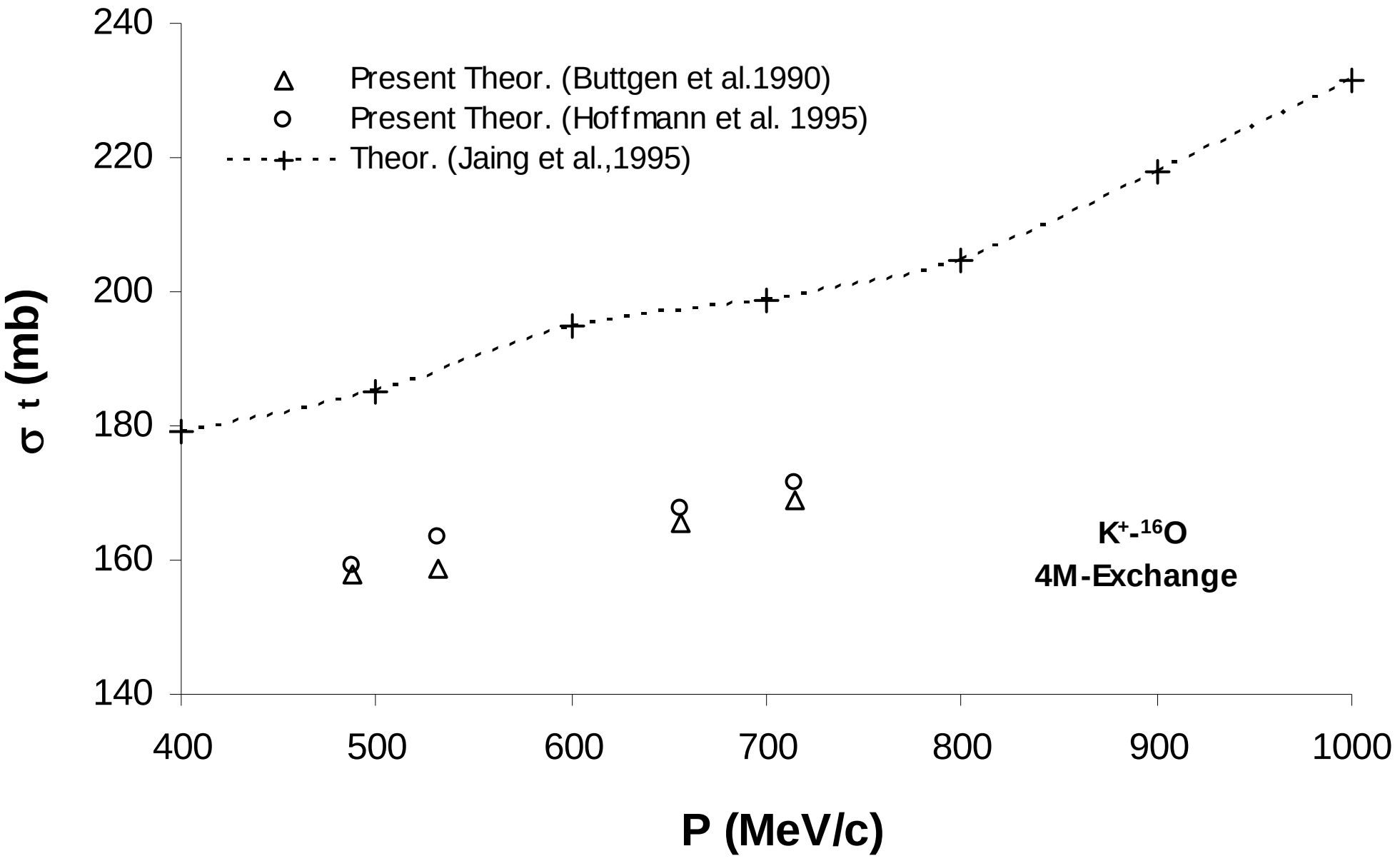

**Fig. (11) The total cross section versus the momentum in $K^{+}-{}^{16}O$ reaction. The theoretical calculations by different parameters used as Buttgen et al. [20] and Hoffmann et al.,[28] respectively.**

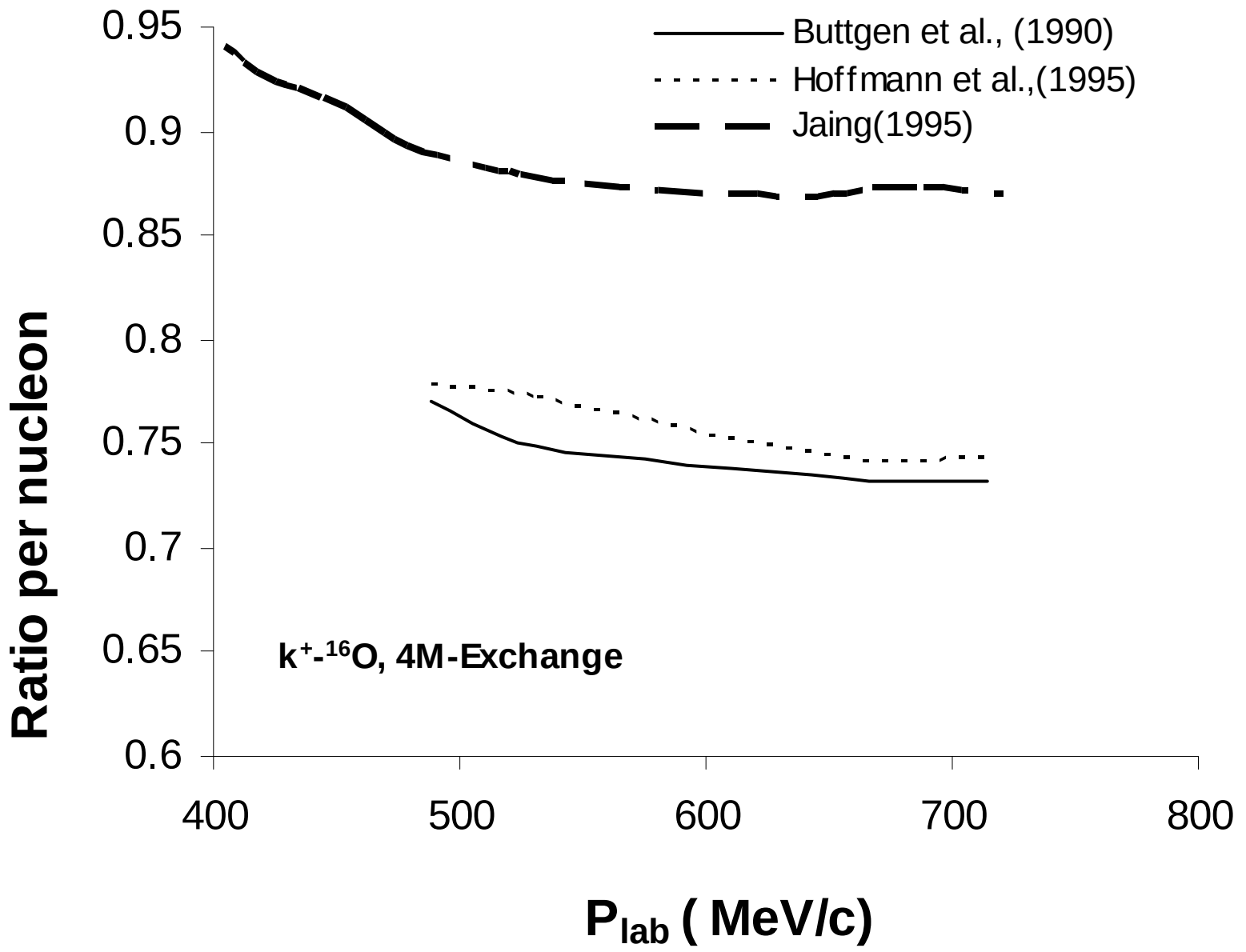

**Fig. (12) The Ratio of Oxygen to Deuterium cross sections per nucleon. The solid line and the dashed line indicate the theoretical calculations by different parameters used as Buttgen et al. [20] and Hoffmann et al.,[28] respectively.**

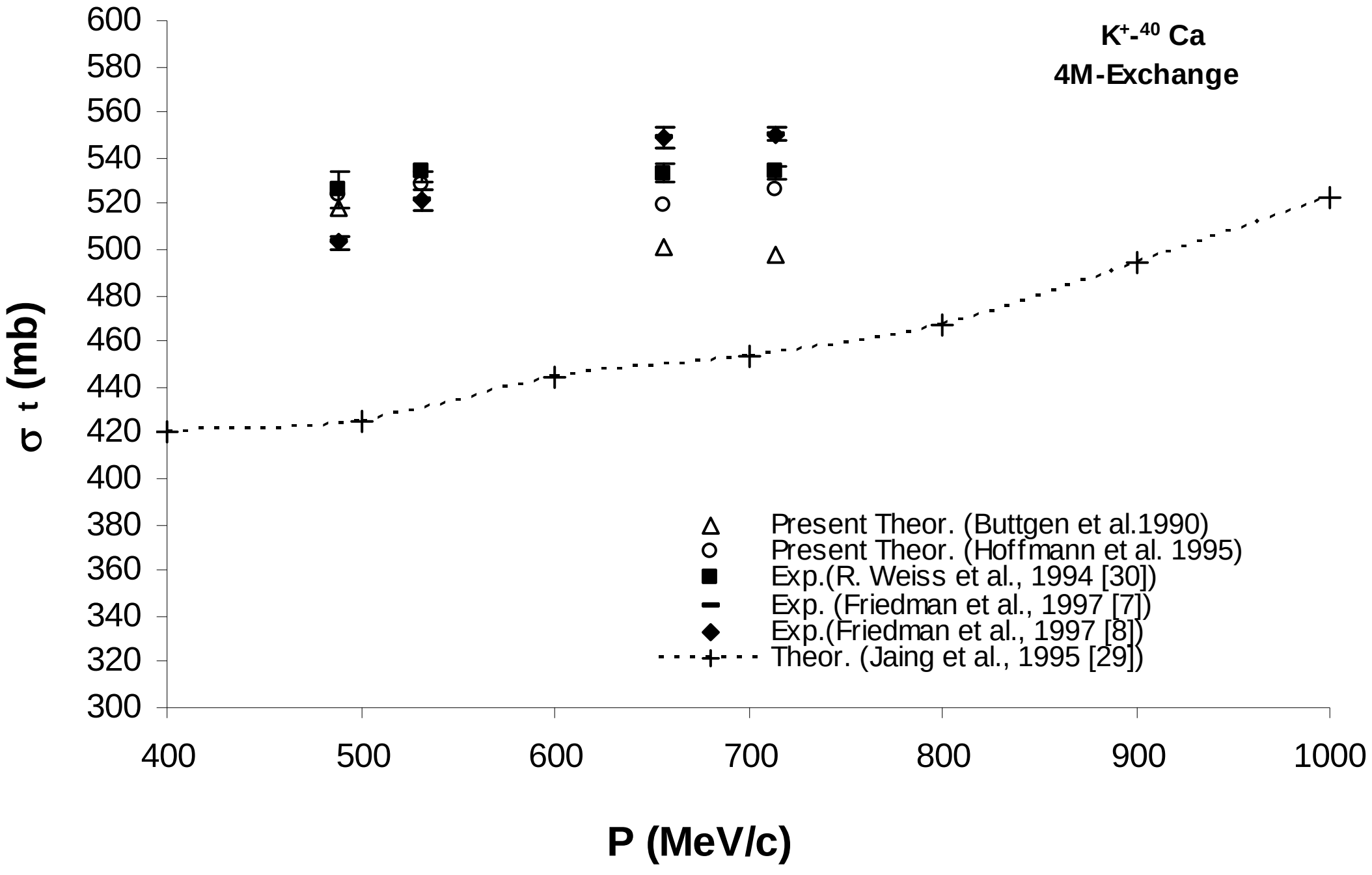

**Fig. (13) The total cross section versus the momentum in $K^{+} - ^{40}Ca$ reaction. The theoretical calculations by different parameters used as Buttgen et al. [20] and Hoffmann et al.,[28] respectively.**

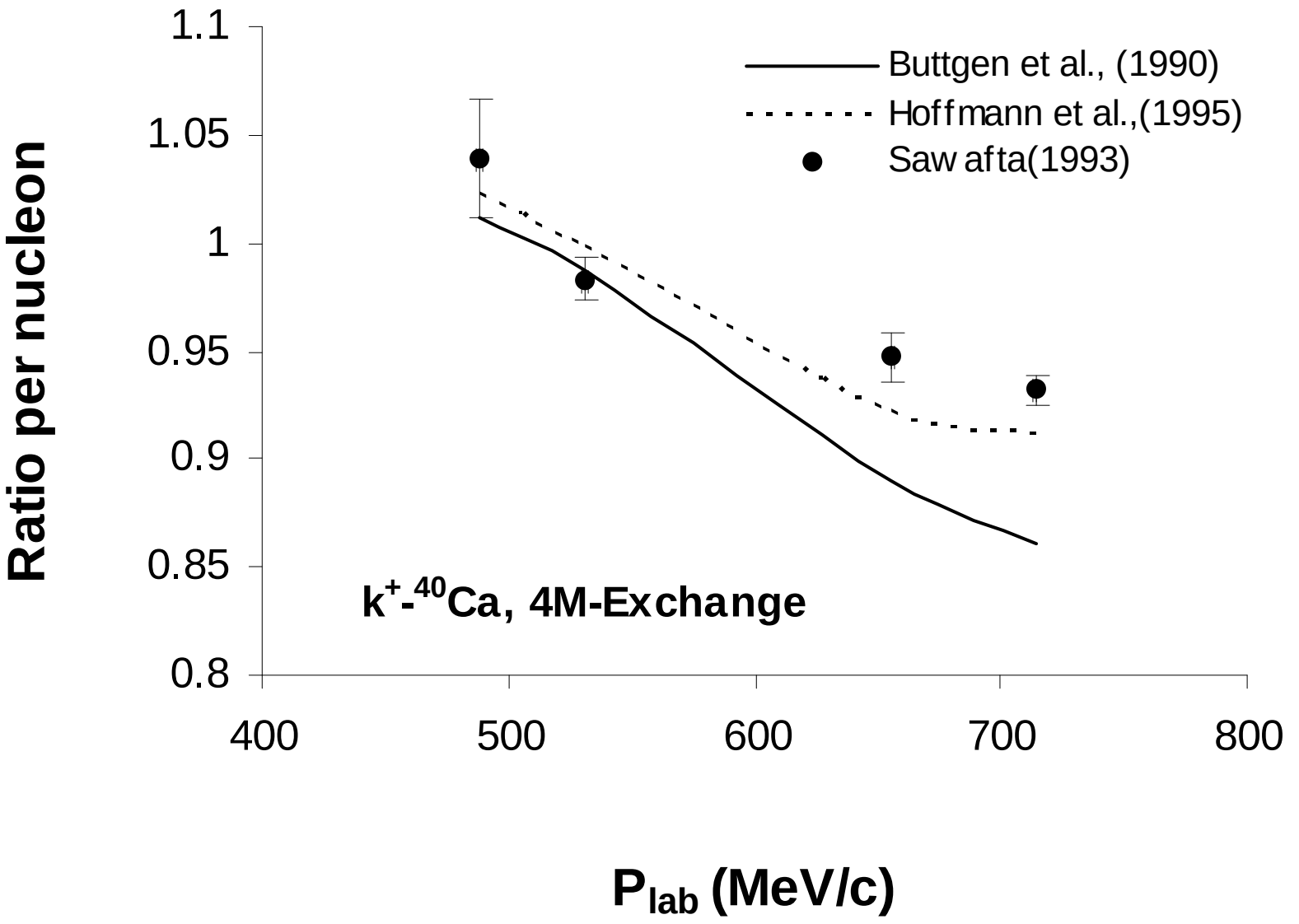

**Fig. (14) The Ratio of Calcium to Deuterium cross sections per nucleon. The solid line and the dashed line indicate the theoretical calculations by different parameters used as Buttgen et al. [20] and Hoffmann et al.,[28] respectively.**

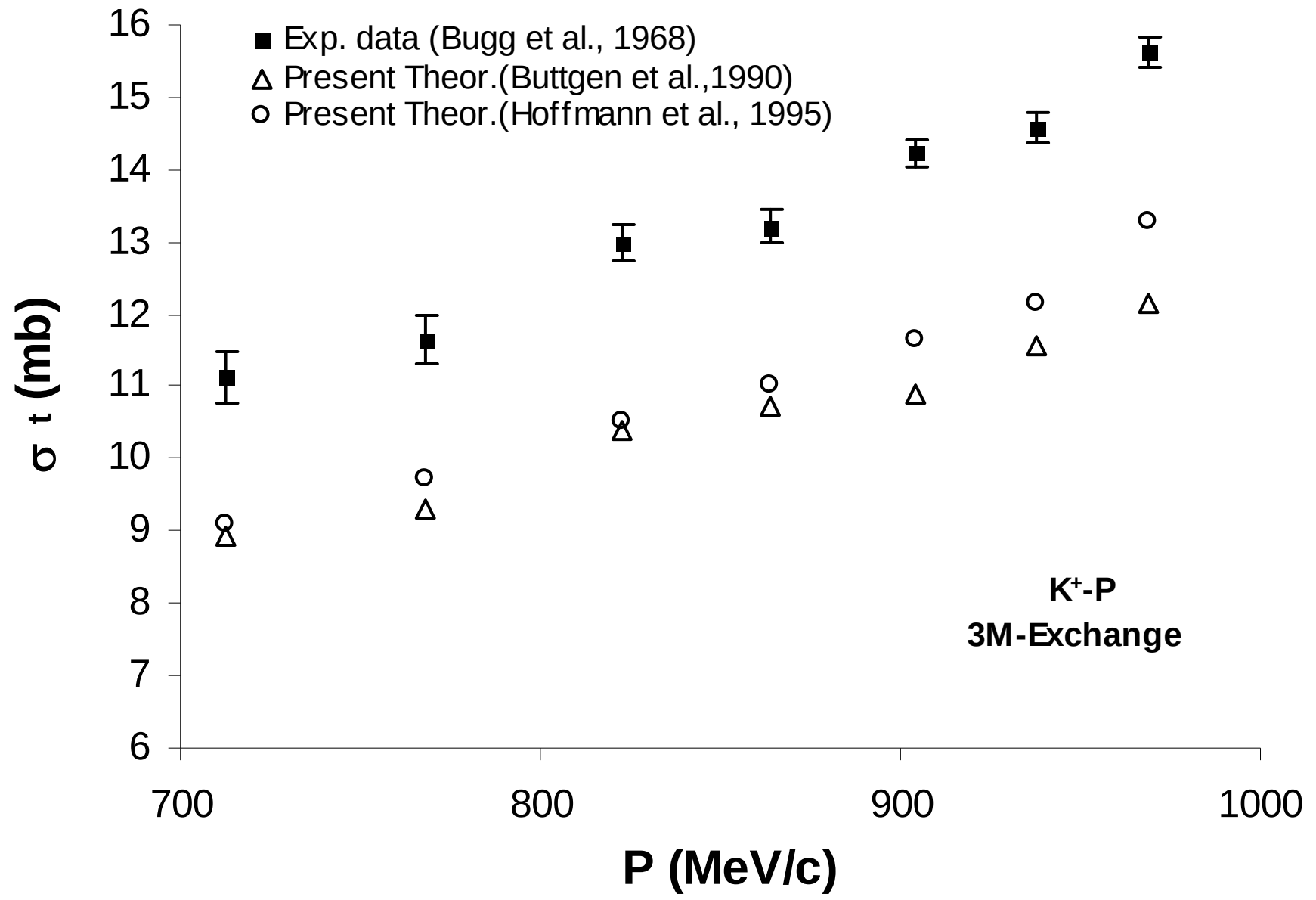

**Fig. (15) The total cross section versus the momentum in $K^+ - P$ reaction. The theoretical calculations by different parameters used as Buttgen et al. [20] and Hoffmann et al.,[28] respectively.**

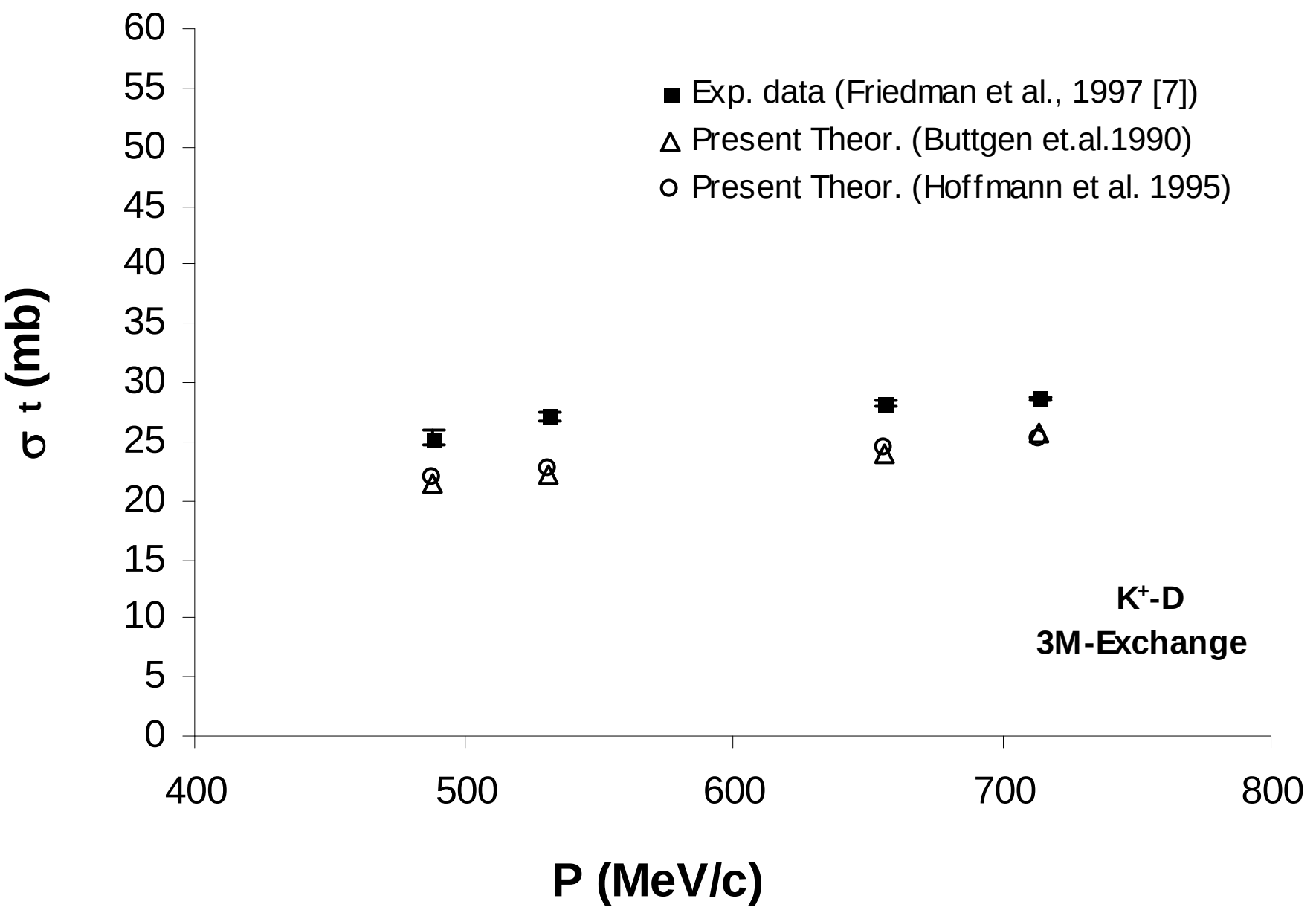

**Fig. (16) The total cross section versus the momentum in $K^+ - Deuteron$ reaction. The theoretical calculations by different parameters used as Buttgen et al. [20] and Hoffmann et al.,[28] respectively.**

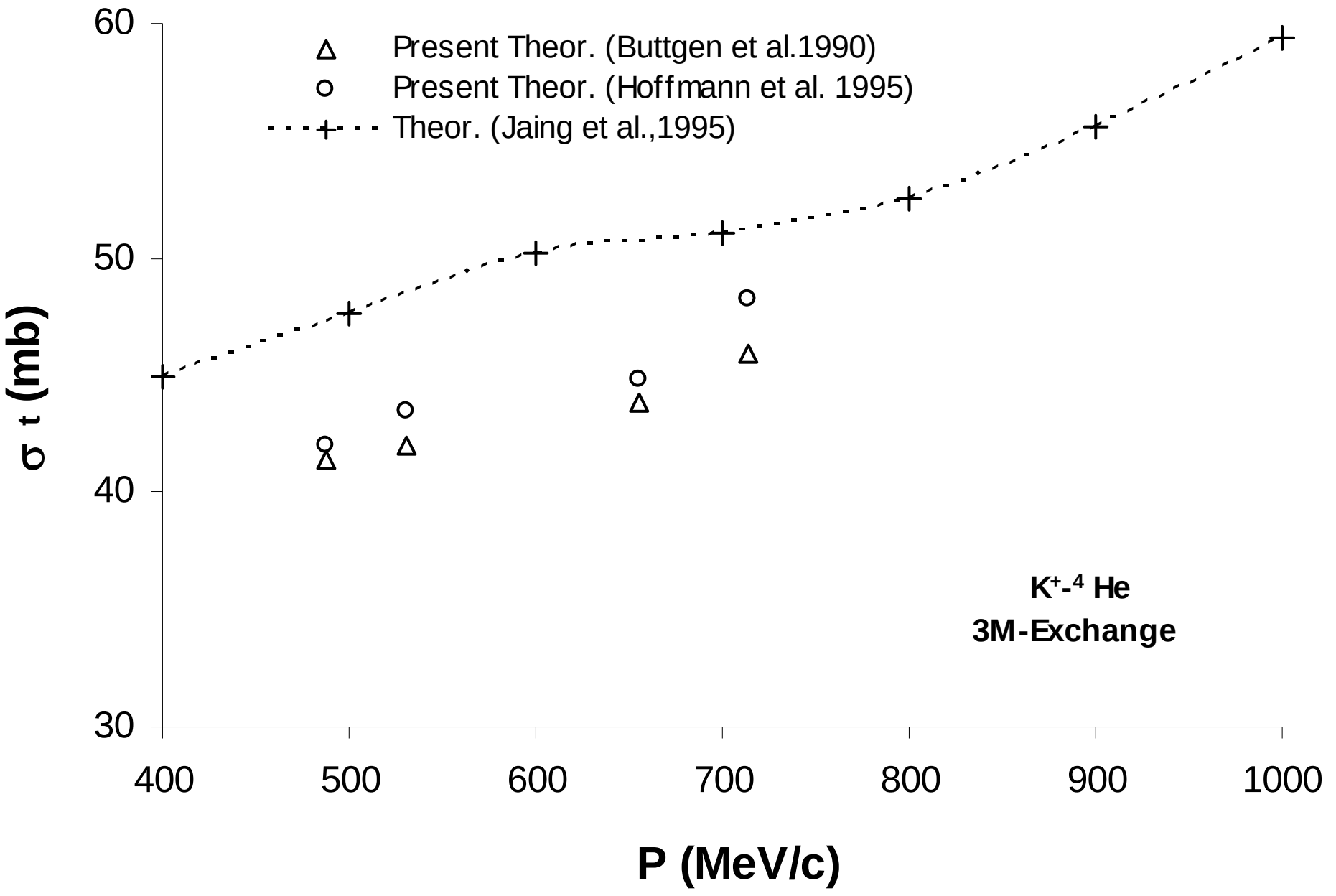

**Fig. (17) The total cross section versus the momentum in $K^{+} - ^{4}He$ reaction. The theoretical calculations by different parameters used as Buttgen et al. [20] and Hoffmann et al.,[28] respectively.**

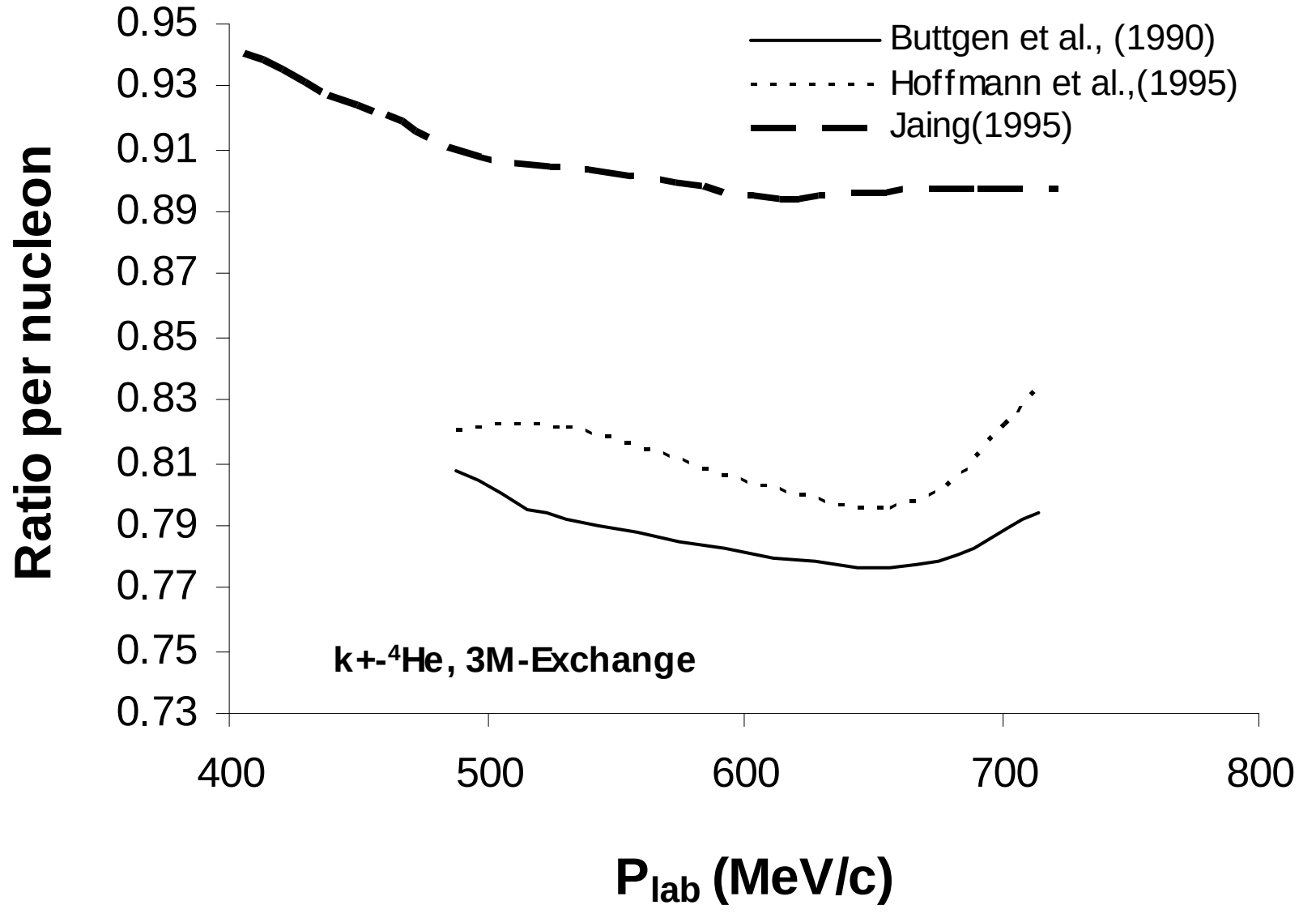

**Fig. (18) The Ratio of Helium to Deuterium cross sections per nucleon. The solid line and the dashed line indicate the theoretical calculations by different parameters used as Buttgen et al. [20] and Hoffmann et al.,[28] respectively.**

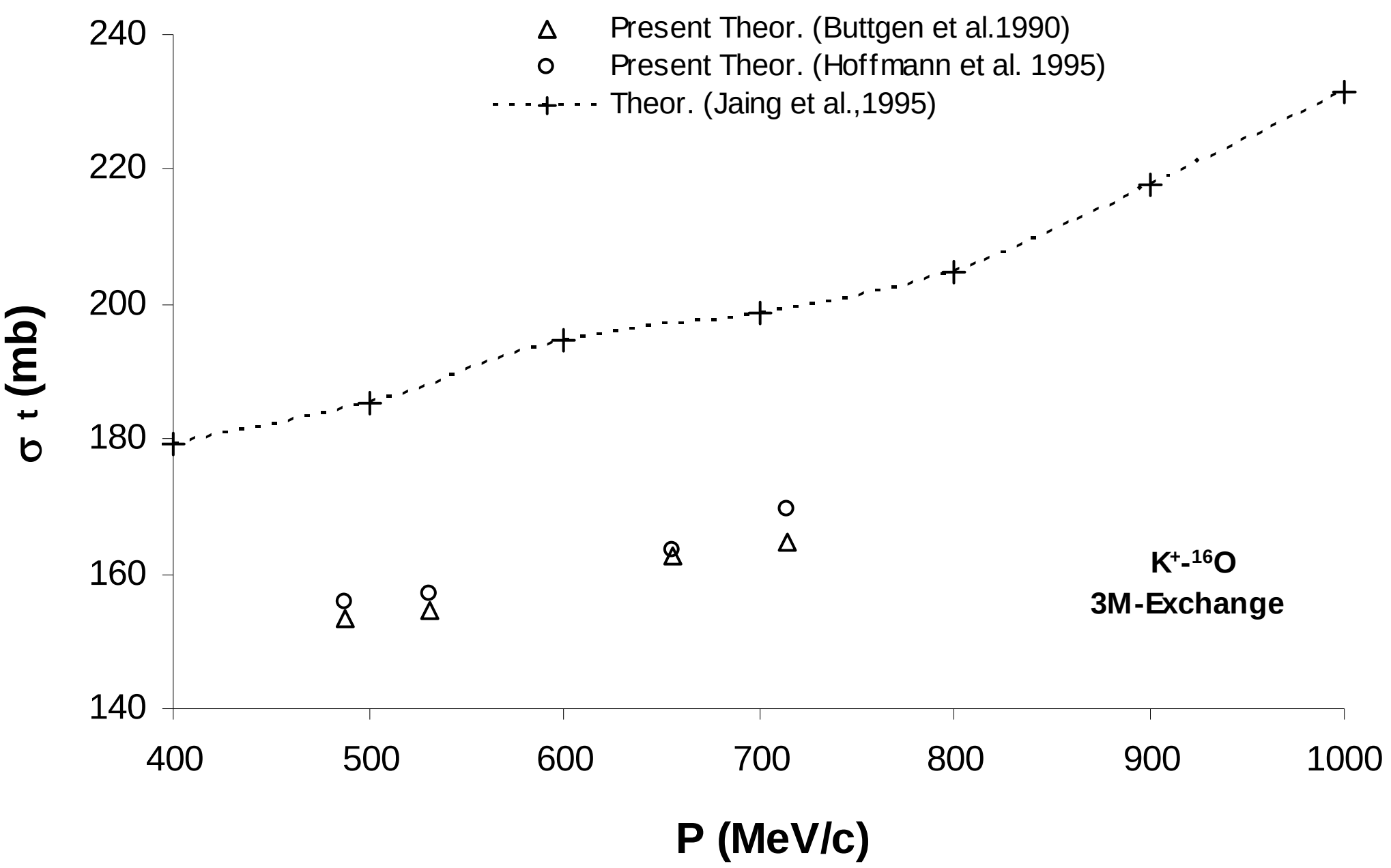

**Fig. (19) The total cross section versus the momentum in $K^{+} -^{16} O$ reaction. The theoretical calculations by different parameters used as Buttgen et al. [20] and Hoffmann et al.,[28] respectively.**

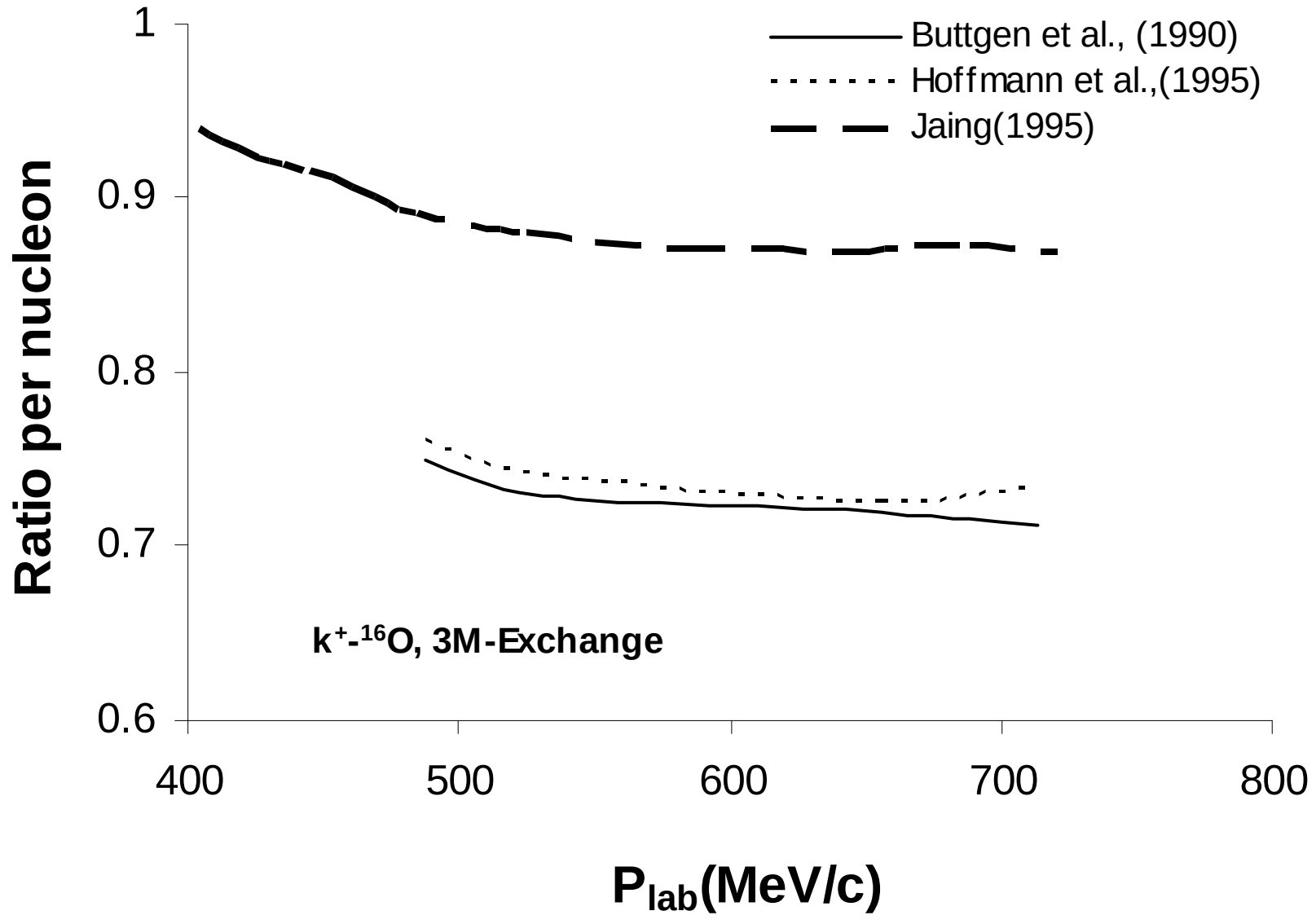

**Fig. (20) The Ratio of Oxygen to Deuterium cross sections per nucleon. The solid line and the dashed line indicate the theoretical calculations by different parameters used as Buttgen et al. [20] and Hoffmann et al.,[28] respectively.**

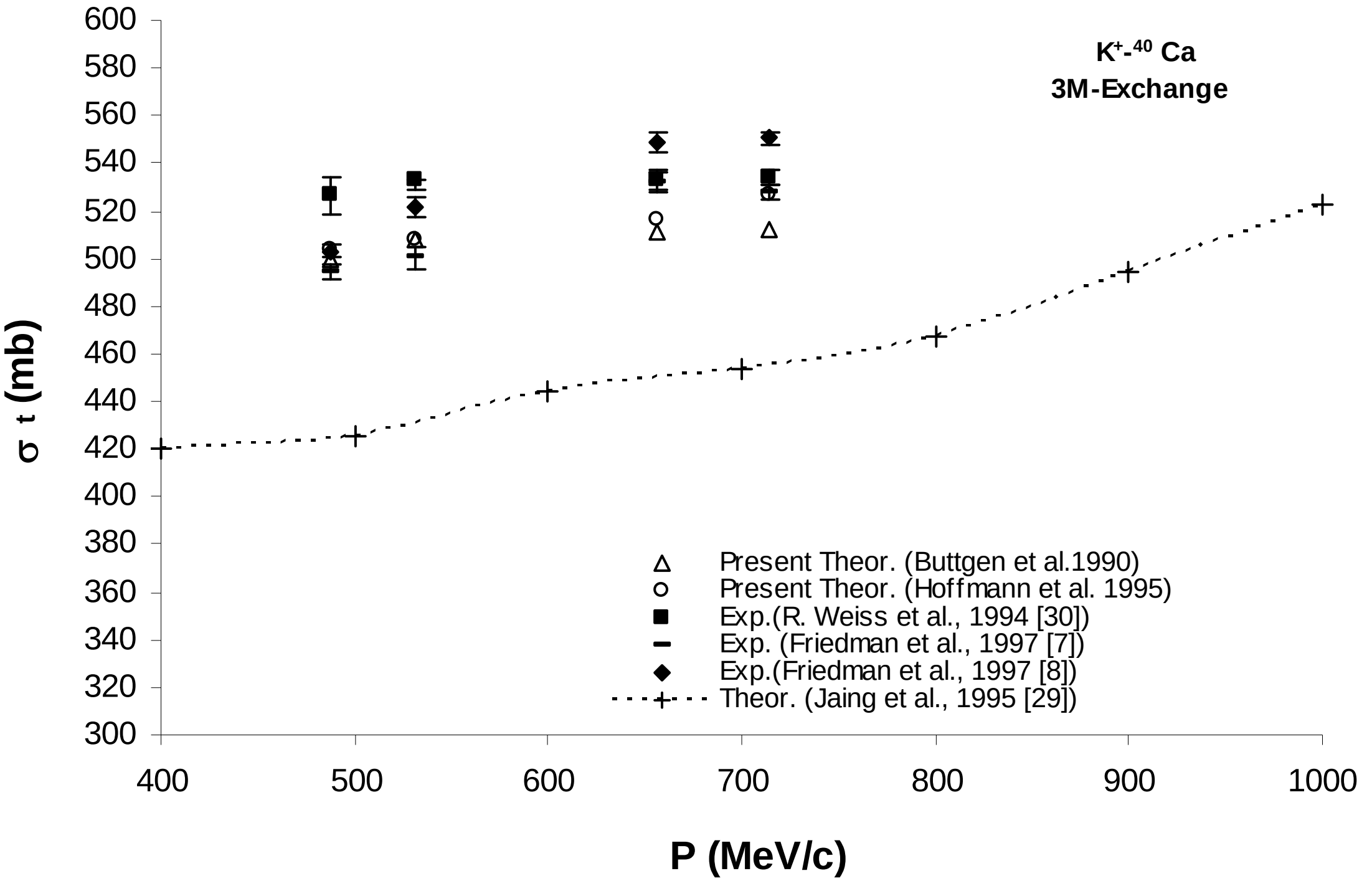

**Fig. (21) The total cross section versus the momentum in $K^{+} - {}^{40}Ca$ reaction. The theoretical calculations by different parameters used as Buttgen et al. [20] and Hoffmann et al.,[28] respectively.**

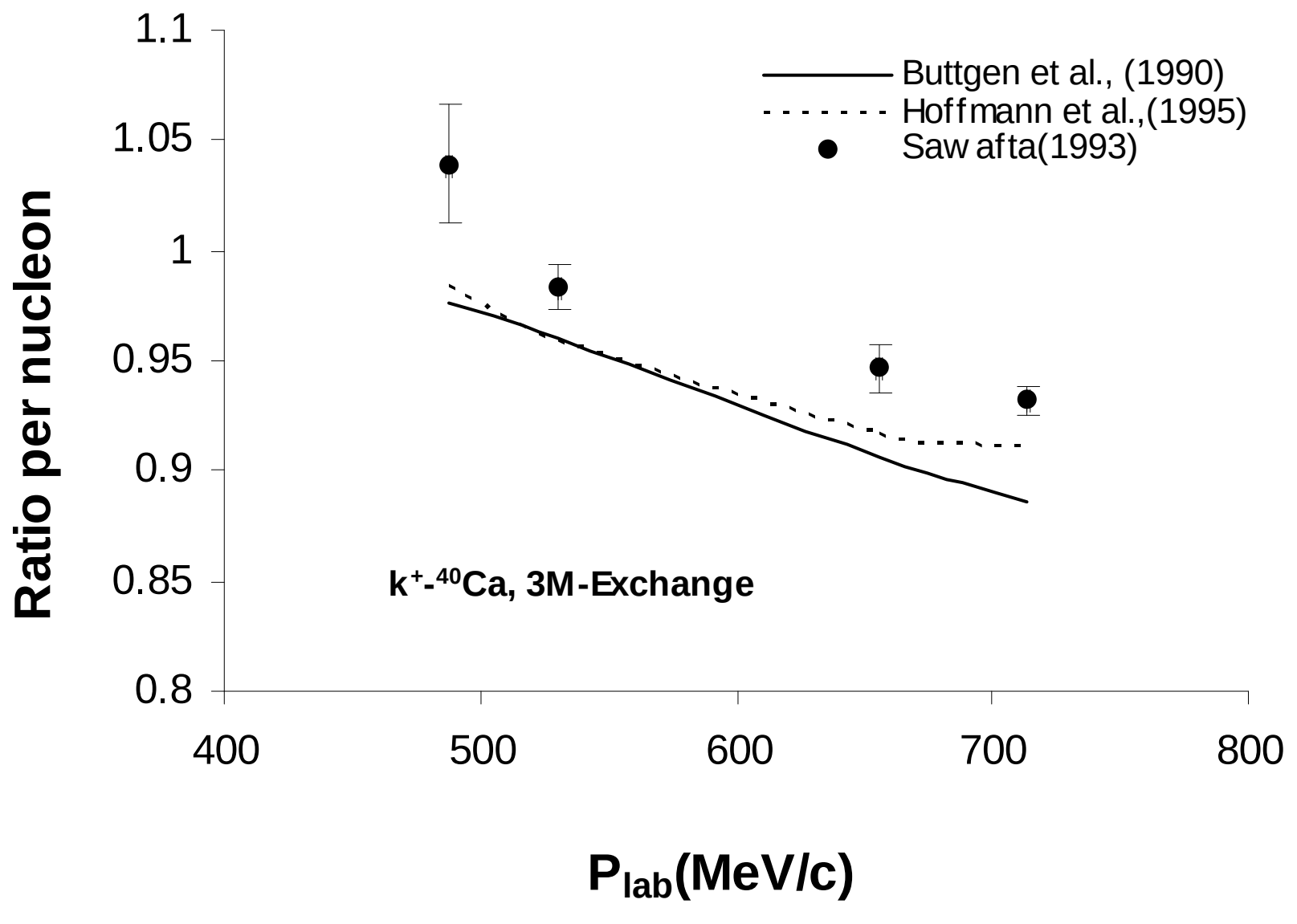

**Fig. (22) The Ratio of Calcium to Deuterium cross sections per nucleon. The solid line and the dashed line indicate the theoretical calculations by different parameters used as Buttgen et al. [20] and Hoffmann et al.,[28] respectively.**

## IV. Conclusion

In the present work the total scattering of the positive Kaon ($K^+$)-nucleon, $K^+$ -$^4$He , $K^+$-$^{16}$O and $K^+$-$^{40}$Ca has been studied at intermediate energy range of the incident Kaon. Important conclusions can be derived from this study as follows,

1. The microscopically derived optical potential used in our calculations, which based on the one-boson-exchange method to describe the reaction between the Kaon-reactant, proved itself a very efficient potential.
2. The potential we derived and applied have two essential features, the repulsive and the short range characters. These features are required and consistent the experimental data.
3. The above two features of the potential encourage us to suffice in our calculations with the impulse approximation, which showed reasonable agreement with the data at the appointed energies.
4. The nuclear medium effects and the charge dependence were reported in our calculations via the wave vector and the Clebsch-Gordon coefficients. The charge dependence has been accounted where in our calculations the interaction of the Kaon with the two different isotopic spin forms of the nucleon give different values for the Clebsch-Gordon coefficients.
5. The spin-orbit force was included in the calculations without using any further adjustable parameters.
6. Due to that the data needs more repulsion and the $\omega$ meson can not bear alone the burden of that, we suggest the use of the $\sigma_0$ meson with heavier mass than the sigma meson and opposite sign to account for the additional repulsion required by the data. In fact, as expected, the calculated results gave more consistency with the measured data.

All what we said above urge us to continue to get more accurate theory approach to investigate the $K^+$, the most weak probe in the strong interaction theory in the near future.